\documentclass{emulateapj}
\usepackage{times}

\newcommand{\boltz}{{\tt BOLTZTRAN}}
\newcommand{\agile}{{\it Agile}}
\newcommand{\aboltz}{\agile-\boltz}

\newcommand{\msun}{\ensuremath{M_\sun}}

\newcommand{\nue}{\ensuremath{\nu_{e}}}
\newcommand{\nuebar}{\ensuremath{\bar \nu_e}}
\newcommand{\numt}{\ensuremath{\nu_{\mu\tau}}}
\newcommand{\numtbar}{\ensuremath{\bar \nu_{\mu\tau}}}
\newcommand{\numu}{\ensuremath{\nu_{\mu}}}
\newcommand{\nutau}{\ensuremath{\nu_{\tau}}}
\newcommand{\numubar}{\ensuremath{\bar \nu_{\mu}}}
\newcommand{\nutaubar}{\ensuremath{\bar \nu_{\tau}}}
\newcommand{\mev}{\mbox{MeV}}
\newcommand{\Ynu}{\ensuremath{Y_{\nu}}}
\newcommand{\Mshock}{\ensuremath{M_{\rm sh}}}
\newcommand{\meanE}[1]{\ensuremath{\langle E_{#1}\rangle_{\rm RMS}}}

\newcommand{\isotope}[2]{$^{#2}$#1}
\newcommand{\gcc}{\ensuremath{{\mbox{g~cm}}^{-3}}}
\newcommand{\den}[2]{\ensuremath{#1 \times 10^{#2}\,\textrm{g~cm}^{-3}}}

\newcommand{\Bethes}{\ensuremath{{\mbox{Bethe~s}}^{-1}}}
\newcommand{\Rnu}[1]{\ensuremath{R_{#1}}}

\newcommand{\base}{Base}
\newcommand{\ipa}{IPA}
\newcommand{\basenis}{\base-noNIS}
\newcommand{\basenes}{\base-noNES}
\newcommand{\basenps}{\base-noNPS}
\newcommand{\basereds}{\base-ISnp}

\newcommand{\ipanis}{\ipa-noNIS}
\newcommand{\ipanes}{\ipa-noNES}
\newcommand{\ipanps}{\ipa-noNPS}
\newcommand{\ipareds}{\ipa-ISnp}

\newcommand{\baserea}{\base-B85ea-np}

\newcommand{\basepair}{\base-noEPpair}
\newcommand{\basenopair}{\base-noPair}
\newcommand{\basebrem}{\base-noBrems}

\shorttitle{Interplay of Neutrino Opacities in Core-collapse Supernovae}
\shortauthors{Lentz et al.}

\slugcomment{Accepted for publication to ApJ: 2 Oct 2012}
\begin{document}

\title{Interplay of Neutrino Opacities in Core-collapse Supernova Simulations}

\author{
Eric J. {Lentz}\altaffilmark{1,2,3},  
Anthony {Mezzacappa}\altaffilmark{2,1,4},
O.E. Bronson {Messer}\altaffilmark{5,1,4}, 
W. Raphael {Hix}\altaffilmark{2,1}, 
and 
Stephen W. {Bruenn}\altaffilmark{6}
}
\email{elentz@utk.edu}

\altaffiltext{1}{Department of Physics and Astronomy, University of Tennessee, Knoxville, TN 37996-1200, USA}
\altaffiltext{2}{Physics Division, Oak Ridge National Laboratory, P.O. Box 2008, Oak Ridge, TN 37831-6354, USA}
\altaffiltext{3}{Joint Institute for Heavy Ion Research, Oak Ridge National Laboratory, P.O. Box 2008, Oak Ridge, TN 37831-6374, USA}
\altaffiltext{4}{Computer Science and Mathematics Division, Oak Ridge National Laboratory, P.O.Box 2008, Oak Ridge, TN 37831-6164, USA}
\altaffiltext{5}{National Center for Computational Sciences, Oak Ridge National Laboratory, P.O. Box 2008, Oak Ridge, TN 37831-6164, USA}
\altaffiltext{6}{Department of Physics, Florida Atlantic University, 777 Glades Road, Boca Raton, FL 33431-0991, USA}

\begin{abstract}

We have conducted a series of numerical experiments using spherically symmetric, general relativistic, neutrino radiation hydrodynamics with the code \aboltz\   to examine the effects of modern neutrino opacities on the development of supernova simulations.
We test the effects of opacities by removing opacities or  by undoing opacity improvements for individual opacities and groups of opacities.
We find that  improvements to electron capture (EC) on nuclei, namely EC on an ensemble of nuclei using modern nuclear structure models rather than  the simpler independent-particle approximation (IPA) for EC on a mean nucleus, plays the most important role during core collapse of all tested neutrino opacities.
Low-energy neutrinos emitted by modern nuclear EC preferentially escape during collapse  without the energy downscattering on electrons required to enhance neutrino escape and deleptonization for the models with IPA nuclear EC.
During shock breakout the primary influence on the emergent neutrinos arises from NIS on electrons.
For the  accretion phase, non-isoenergetic scattering on free nucleons and pair emission by $e^+e^-$ annihilation have the largest impact on the neutrino emission and shock evolution.
Other opacities evaluated, including nucleon--nucleon bremsstrahlung and especially neutrino--positron scattering, have little measurable impact on neutrino emission or shock dynamics.
Modern treatments of nuclear electron capture, $e^+e^-$-annihilation pair emission, and non-isoenergetic scattering on electrons and free nucleons are critical elements of core-collapse simulations of all dimensionality.
\end{abstract}


\section{Introduction}

As the core of a massive star collapses, electron capture (EC) on protons (free or within nuclei) reduces the electron fraction, $Y_e$,  and releases neutrinos that stream from the core.
At densities near \den{6}{11}, the mean-free path for neutrinos of mean energy becomes comparable to the size of the core and the neutrinos become ``trapped.''
The trapping of neutrinos effectively halts the deleptonization (reduction of $Y_e$) of the core as the emission of neutrinos through EC is balanced by reabsorption of the neutrinos on neutrons (either free or within nuclei).
Neutrino opacity regulates the deleptonization and sets the minimum core $Y_L$ ($Y_e$ plus the neutrino fraction, $Y_\nu$),  which controls the size of the homologous inner core, $\Mshock \propto {Y_L}^2$ \citep{Yahi83}.
When the density in the core exceeds nuclear density, the nuclear equation of state (EoS) stiffens, and a bounce shock forms at the sonic point that defines the edge of the homologous core.
The expanding shock loses energy to neutrino emission and nuclear dissociation and stalls.
The neutrino weak interactions (emission, absorption, and  scattering) regulate energy loss by the prompt shock to neutrino radiation and thus affect the stalling of the shock.

The revival of the stalled shock by neutrino heating is likewise regulated by  the neutrino opacities.
As for photons in stellar atmospheres, we can define ``neutrinospheres,'' where the neutrinos of each species and energy effectively decouple from the matter in the stellar core.
The neutrinosphere radii, \Rnu{\nu}, are dependent on both flavor and energy with \Rnu{\nue} larger than \Rnu{\numt} because of the  increased number of weak interaction channels for \nue\ (neutral and charged current).
The neutrinosphere radii increase with increasing neutrino energy due to the larger opacities for higher energy neutrinos.
The emission temperatures at the various neutrinospheres effectively set the flux of neutrinos in each energy--species group and therefore the spectrum of neutrinos available for absorption in the semi-transparent heating, or ``gain,'' region between the proto-neutron star and the shock.

Non-isoenergetic scattering (NIS)  changes the energy and direction of the neutrinos, each of which plays an important role in all phases of the supernova evolution.
Scattering to a lower energy can change the neutrino's environment from opaque to semi-transparent and allow a neutrino that would otherwise be trapped to escape, both increasing the luminosity of neutrinos available for absorption in the gain region above the core and enhancing lepton escape from the core.

The earliest models of neutrino heating in core-collapse supernovae by \citet{CoWh66} included only emission by electron capture and redeposited  half of the energy lost to core neutrino emission in the layers  above the proto-neutron star, driving a powerful explosion of the outer layers of the star.
Subsequent spherically symmetric models included more detailed treatment of neutrino transport to revive the shock: gray diffusion \citep{Arne66}, two-fluid schemes \citep{HiNoWo84,CovaBa86}, multi-group flux-limited diffusion \citep{Brue75,Arne77,BoWi82,Brue85,MyBlHo87}, and Boltzmann transport \citep{MeBr93b,YaJaSu99}.

\begin{deluxetable*}{cccc}
\tabletypesize{\scriptsize}
\tablecaption{Neutrino Opacity Summary Table\label{tab:opac}}
\tablecolumns{4}
\tablewidth{0pt}
\tablehead{
\colhead{Interaction} & \colhead{Base} & \colhead{Alternate}  & \colhead{Model}
}
\startdata
$\nu e^- \leftrightarrow \nu' e^-$   & \citet{SchSh82} & None & \basenes \\
$\nu e^+ \leftrightarrow \nu' e^+$ & &  & \basenps \\
$\nu n \leftrightarrow \nu' n$          & \citet{RePrLa98} & \citet{Brue85} & \basereds  \\
$\nu p \leftrightarrow \nu' p$          &&  &\\
$e^- p \leftrightarrow \nu_e n $     &\citet{RePrLa98} & \citet{Brue85} & \baserea \\
 $e^+ n \leftrightarrow \bar{\nu}_e p $ &&  & \\
 $\nu A \leftrightarrow \nu A$              & \citet{Brue85} & \emph{No Change} & \nodata \\
$\nu \alpha \leftrightarrow \nu \alpha$ & \citet{Brue85} & \emph{No Change} & \nodata \\
$e^- (A,Z) \leftrightarrow \nu_e (A,Z-1) $  &\citet{LaMa00}, & \citet{Brue85}, & \ipa \\
                                                                   &\citet{LaMaSa03} & \cite{Full82}  & \\
$ e^+e^- \leftrightarrow \nu \bar{\nu} $ & \citet{SchSh82}  &None & \basepair \\
$ NN \leftrightarrow NN\nu\bar{\nu}$&  \citet{HaRa98}   & None & \basebrem 
\enddata
\end{deluxetable*}

The neutrino opacities used evolved in complexity and completeness \citep{TuSch75,LaPe76,YuBu76,BlvR78} using the new weak interaction theory \citep{Glas61,Wein67,Sala68} concurrently with relevant experimental results including: the detection of neutral-current interactions in neutrino--nucleon scattering \citep{Haetal73} and the discovery of  $\tau^-$ \citep{Perl75}, whose neutrino partner, \nutau, was widely assumed to exist, but was not detected until the turn-of-the-millenium \citep{DONUT01}.

\citet{SchSh82} and \citet{Brue85} assembled  comprehensive  neutrino opacity sets that included energy coupling in the scattering (NIS) and  were oriented toward numerical implementation.
The  \citet{Brue85} opacity set (``B85'' hereafter) has been widely adapted both in content and form and has widely served as the ``canonical'' opacity set.
The B85 opacity set includes emission, absorption, and isoenergetic scattering (IS) on heavy nuclei, $\alpha$-particles, and free nucleons;  NIS on electrons; and $\nu\bar{\nu}$-pair emission from $e^+e^-$ annihilation. (Though derived in \cite{Brue85}, NIS on positrons is often omitted from simulations using B85 opacities.)

In the more than 25 years since the compilation of the B85 opacity set, the search for opacity improvements (driven in part, previously, by the possibility that spherically symmetric models might explode if sufficiently detailed) has continued.
Some of the newer neutrino channels identified, developed for simulations, and widely adopted include: 
nucleon--nucleon bremsstrahlung \citep[][and a related inelastic $\nu$-nucleon scattering channel]{HaRa98};
more kinematically complete $\nu$-nucleon emission, absorption, and scattering opacities  \citep{BuSa98,RePrLa98};
the pair--flavor conversion process \citep{BuJaKe03};
and tabulated EC rates using ensembles of nuclei with detailed level structures \citep{LaMaSa03,JuLaHi10}.
Other refinements and corrections to  opacities that have been added to many simulations include:
weak magnetism for interactions with free nucleons \citep{Horo02};
ion-ion correlations between nuclei \citep{BoWi82,Horo97,ItAsTo04};
and changes from the effective mass of the nucleons in dense matter \citep[cf.,][]{RePrLa99}.
For dense matter, from where nuclei become correlated through nuclear matter, the next frontier in the computation of $\nu$--nucleon/nucleus interactions is the development of  EoS tables with consistent neutrino opacities for emission, absorption, scattering, and neutrino-pair processes \citep{RePrLa98,MaFiLo12,RoRe12}.

Updated neutrino opacity sets have been assembled by several authors \citep{Burr01,BuRaJa06,BrDiMe06,LeMeMe12}.
Prior studies in spherical symmetry have examined the impact of the addition, or modification, of a single opacity \citep{MeBr93c,BuJaKe03,MeLiHi03,HiMeMe03,MaJaBu05,LaMaMu08} and of multiple, simultaneous, opacity changes in spherical symmetry \citep{BuRaJa06,LeMeMe12} and in axisymmetry (2D) \citep{BuRaJa06,MuJaMa12}.
This is the first study detailing the effects of each of the opacity changes made to create a modernized opacity set.
In this paper, we start from the full set of opacities used in \citet{LeMeMe12} and test not only each component of the opacity change in that paper and each opacity upgrade relative to the B85 opacities, but also reexamine the omission of neutrino--electron scattering (NES) and all NIS opacities and omission of each pair-source opacity.
In each of our simulations we retain at minimum the B85-formulation of scattering on nucleons and nuclei and emission and absorption on nucleons and nuclei to ensure that the total opacity is not radically changed.
While the differences found among the tests of opacity removal are generally consistent with prior single-opacity studies, we find that opacity changes in the context of a detailed opacity set can have different impacts than changing the same opacity in a less complete opacity set.
This contextual effect is most prominent for NES during collapse, where the previously identified role of NES in enhancing neutrino escape and core deleptonization during collapse by downscattering neutrinos to lower energies is muted by detailed EC on nuclei.
We identify, within our modern opacity set, critical opacities needed for reliable computation of the shock dynamics and neutrino emission during the collapse, shock breakout, and accretion phases, as well as opacities of little impact on the simulation or the observational neutrino properties.

\section{Numerical methods and inputs}

All models in this paper are computed using the parallel version of \aboltz, a code for general-relativistic, spherically-symmetric, neutrino radiation hydrodynamics \citep{LiMeMe04} with extensions described here.

\begin{deluxetable*}{cccccccc}
\tabletypesize{\scriptsize}
\tablecaption{Model Summary Table\label{tab:models}}
\tablecolumns{8}
\tablewidth{0pt}
\tablehead{
\colhead{} & \multicolumn{4}{c}{Bounce properties} &\colhead{}& \multicolumn{2}{c}{Post-bounce peak} \\
\cline{2-5} \cline{7-8}\\
\colhead{Model} & \colhead{Core mass (\Mshock)} &  \colhead{central $\rho_c$} &  \colhead{central $Y_e$} &  \colhead{central $Y_L$} &  \colhead{} &  \colhead{shock radius} &  \colhead{\nue-luminosity}  \\
\colhead{}            & \colhead{\msun} &  \colhead{$ 10^{14}\, \gcc$} &  \colhead{} &  \colhead{} &  \colhead{} &  \colhead{km} &  \colhead{\Bethes}  
}
\startdata
\base       &  0.430 & 3.234 & 0.2448 & 0.2804 && 161 & 408 \\
\basenis  & 0.431 & 3.234 & 0.2453 & 0.2811 && 150 & 478 \\
\basenes & 0.430 & 3.234 & 0.2450 & 0.2807 && 158 & 481\\
\basereds & 0.431 & 3.233 & 0.2451 & 0.2808 && 153 & 404\\
\basenps  & 0.430 & 3.233 & 0.2448 & 0.2804 && 160 & 408\\
\ipa & 0.554 & 3.824 & 0.2843 & 0.3331 && 159 & 432 \\
\ipanis & 0.618 & 4.239 & 0.3099 & 0.3712 && 148 & 449\\
\ipanes & 0.608 & 4.162 & 0.3056 & 0.3647 && 159 & 454\\
\ipareds & 0.554 & 3.831 & 0.2849 & 0.3339 && 149 &430\\
\ipanps  & 0.551 & 3.825 & 0.2843 & 0.3331 && 159 & 432\\
\basepair & 0.431 & 3.233 & 0.2448 & 0.2804 && 183 & 407 \\
\basebrem & 0.430 & 3.216 & 0.2443 & 0.2798 &&163 & 407 \\
\basenopair & 0.435 & 3.216 & 0.2443 & 0.2798 && 185 & 410 \\
\baserea  & 0.431 & 3.239 & 0.2452 & 0.2808 && 159 & 393
\enddata
\end{deluxetable*}

\begin{figure*}
\epsscale{1.1}
\plotone{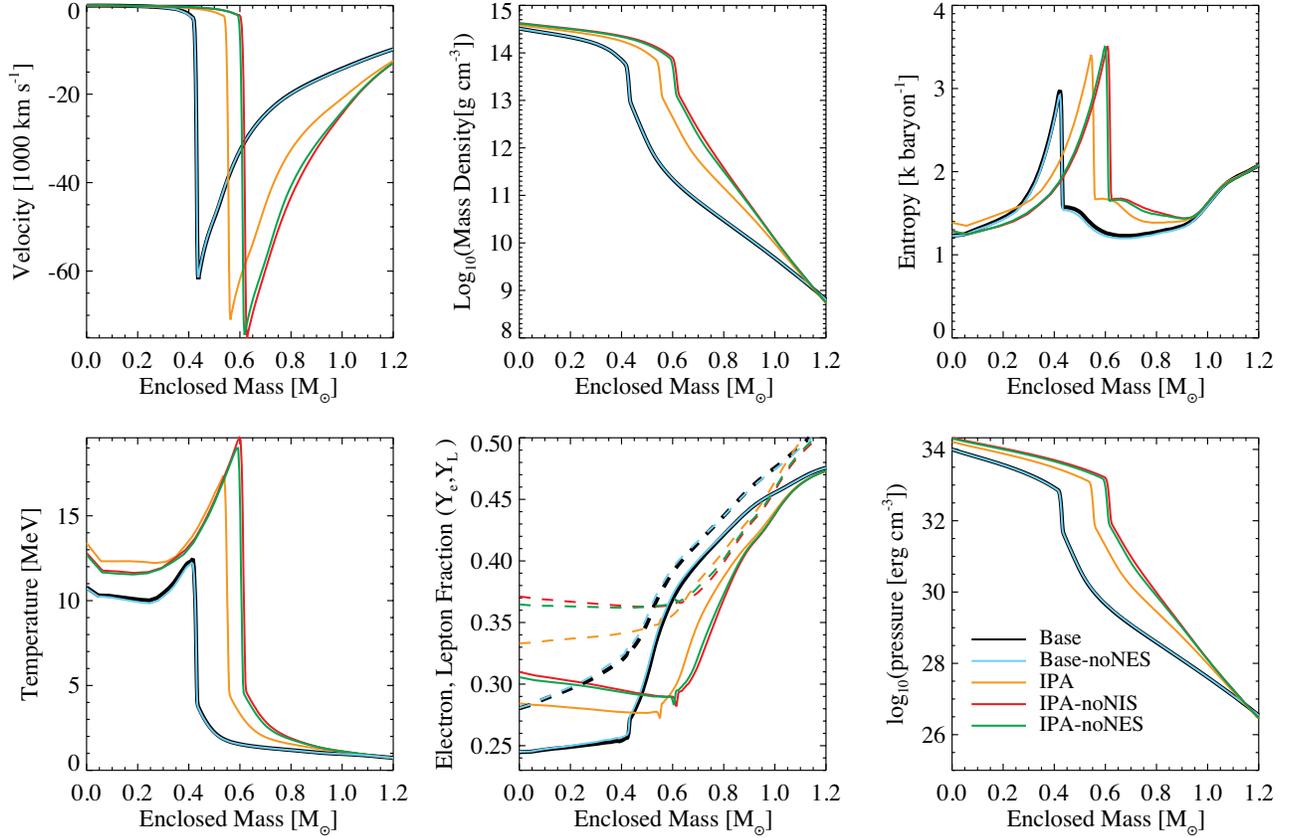}
\caption{Properties of models at core bounce, where bounce is defined as the maximum compression of the central density during the launching of the bounce shock. Models shown are: \base\ (black; all opacities), \basenes\ (blue; without NES) with the other NIS opacity variation models discussed in Section~\ref{sec:basenis} indistinguishable from models \base\ and \basenes\ at bounce and omitted for clarity; the NIS opacity variation models with the IPA EC from Section~\ref{sec:ipanis}, \ipa\ (orange; all NIS opacities), \ipanis\ (red; without NIS opacities; no NES, no NPS, nucleon IS), \ipanes\ (green; without NES), but not  \ipareds\  (nucleon IS), which is indistinguishable from model \ipa\ at bounce and omitted for clarity. The pair opacity test models (Section~\ref{sec:pair}) and improved nucleon EC model (Section~\ref{sec:abem}) are also indistinguishable from model \base\ and omitted for clarity. The panels are: radial velocity (upper left), density (upper center), entropy (upper right), temperature ($kT$, lower left), net electron (or proton) fraction ($Y_e$, lower center, solid lines), net lepton fraction ($Y_L = Y_e +  (n_{\nu_e} - n_{\bar{\nu}_e})/n_{\rm baryons}$, lower center, dashed lines), and pressure (lower right). All quantities are plotted relative to enclosed rest-mass in \msun. \label{fig:bounce}}
\end{figure*}

\subsection{\aboltz}

\aboltz\ is a combination of the general-relativistic (GR) hydrodynamics code \agile\ \citep{LiRoTh02} and the neutrino transport code \boltz\ \citep{MeBr93b,MeMe99,LiMeMe04}.
\agile\ solves the complete GR spacetime and hydrodynamics equations implicitly in spherical symmetry on a dynamic, moving grid.
The moving grid allows us to adequately resolve whole collapsing core, including the shock, using only $\mathcal{O}(100)$ radial zones.
Enhancements include the use of a TVD (total variation diminishing) advection scheme in the hydrodynamics solver \citep{LiRaJa05}, which improves the accuracy of advection, and the use of  $\delta m$ as the grid coordinate rather than the enclosed mass  \citep[][ Section~2.1]{FiWhMe10}, which improves numerical accuracy when mass zones are small and  density gradients are large.
\boltz\ \citep{MeBr93b,MeMe99,LiMeMe04} solves the GR neutrino Boltzmann equation using the method of discrete ordinates ($S_N$) with a  Gauss-Legendre  quadrature.
Here we use an 8-point angular quadrature  and 20 logarithmically-spaced energy groups, with group centers from 3 to 300~\mev.
The discretization scheme is designed to simultaneously conserve lepton number and energy as  described in \citet{LiMeMe04}.
Since we do not include any physics to distinguish between muon- and tau-flavored leptons, we use the combined species $\numt = \{\numu,\nutau\}$ and $\numtbar = \{\numubar,\nutaubar\}$.

\subsection{Opacities and other inputs}

For all models we use the nuclear, electron, and photon equations of state (EoS) of \citet{LaSw91} with the bulk incompressibility of nuclear matter $\kappa_s = 220 \, \mev$.\footnote{We use the latest version of the \citet{LaSw91} EoS, version 2.7, which is available for download from its authors at \url{http://www.astro.sunysb.edu/dswesty/lseos.html}.}
This matches the current experimental value of $\kappa_s = 240 \pm 20 \, \mev$ \citep{ShKoCo06} better than the  value of 180~\mev\ more commonly used with LS~EoS in the past, though the value of $\kappa_s$ has been shown to be of little consequence during the early phases of core-collapse supernova evolution shown here \citep{SwLaMy94,ThBuPi03,LeHiBa10}.
Matter outside the ``iron'' core\footnote{The Fe-core is defined as the inner core that is in nuclear statistical equilibrium (NSE) or where the mass fraction of Fe-peak nuclei exceeds 0.5.} is treated as an ideal gas of \isotope{Si}{28} that  ``flashes'' instantaneously to nuclear statistical equilibrium when the temperature exceeds $kT > 0.47 \, \mev$.

The stellar progenitor used for all models reported here is the 15-\msun\ solar-metalicity progenitor of \citet{WoHe07}. We have mapped the inner $1.8 \, \msun$ of the progenitor onto 108 mass shells of the adaptive radial grid.

The base  opacity set  includes emission, absorption, and scattering on free nucleons \citep{RePrLa98}; isoenergetic scattering on heavy nuclei and $\alpha$-particles \citep{Brue85}; scattering of neutrinos on electrons  (NES) and positrons (NPS)  \citep{SchSh82}; production of neutrino pairs from $e^+e^-$ annihilation \citep{SchSh82} and nucleon-nucleon bremsstrahlung \citep{HaRa98}; and  electron capture (EC) on nuclei using the LMSH EC table of \citet{LaMaSa03}, which utilizes the EC rates of \citet{LaMa00} and \citet{SaLaMa02}.
The full angle and  energy exchange for scattering between the neutrinos and  electrons, positrons, and nucleons is included, while  scattering on nuclei is isoenergetic.
Bremsstrahlung and $e^+e^-$ annihilation are the only sources of \numt\ and \numtbar.

In this paper we conduct numerical experiments where individual opacities in the modernized base set are replaced with  alternatives, or removed.
For NES, NPS, and the $e^+e^-$-annihilation and bremsstrahlung pair sources we have no alternatives so these opacities are tested by removal.
For emission, absorption, and scattering on nucleons, and EC on nuclei, the base opacity components are replaced individually by the simpler versions in \citet{Brue85}.
The alternative to the LMSH EC table, which includes detailed emission rates over an ensemble of nuclei, is an independent particle approximation (IPA)  \citep{Full82,Brue85}, which cuts off when the mean neutron number of the heavy nuclei  $N \geq 40$.
The alternative to the NIS nucleon scattering opacities of \citet{RePrLa98} are the more approximate IS equivalents from \citet{Brue85}, which  include the phase space, but not the recoil of the nucleons.
The alternative to the neutrino emission and absorption on free nucleon opacities of \citet{RePrLa98} from \citet{Brue85} uses an approximate phase space factor and omits nucleon recoil effects.
Ion-ion correlations and weak magnetism are omitted from our opacity set and tests.
The scattering on nuclei remains the \citet{Brue85} form for all models.
The base opacities and their alternatives along with the models testing each alternative are summarized in Table~\ref{tab:opac}.

\section{Results}

A previous paper \citep{LeMeMe12} examined the effects of removing significant sections of the available modern opacity set with noticeable consequences.
However, the effects of individual opacity changes were not isolated and the opacity comparison was conducted using Newtonian hydrodynamics and gravity and $\mathcal{O}(v/c)$ transport.
In this paper we start from the same general relativistic model with the full modern opacities and examine the effects of the various opacities in detail.
The tests are organized into three groups: 1) non-isoenergetic scattering opacity tests, 2) emission/absorption opacity tests, and 3)  pair opacity tests.
Each modern opacity is removed or replaced with an alternative, individually, with additional  models in which groups of opacities are removed or replaced with alternatives.
The models are summarized in Table~\ref{tab:models}.
The configurations at bounce of all distinguishable models are plotted in Figure~\ref{fig:bounce}.
We limit our simulations to the first 150~ms after bounce when shock radius, proto-neutron star radius, and thermodynamic profiles are reasonable approximations to multi-dimensional models.

\subsection{Non-isoenergetic Scattering Comparisons}

There is a rich literature \citep{Brue85,BrHa91,MeBr93c,SmCevdH96,ThBuPi03} on NES effects in core-collapse
demonstrating that, despite being a relatively small contributor to the total scattering opacity,  neutrino--electron scattering (NES) plays an important role in determining core deleptonization during collapse.
The dominant opacities that control the flow of neutrinos and lepton number from the core are energy dependent, with larger opacities (lower mean free paths) for higher neutrino energies.
Energy downscattering by neutrinos on electrons (or other constituents) lowers the energy and therefore the optical depth of the scattered neutrino.
This opens a natural channel for enhanced neutrino escape that is artificially suppressed when NIS opacities are excluded from the core-collapse model.

We retest the effects of energy downscattering on collapse and the subsequent evolution of the collapsed core by removing each source of NIS  individually from our reference model \base.
We also compare to a model that includes no NIS opacities.
Surprisingly, given the previous studies,  none of these changes affected the collapse phase, though differences do emerge after core bounce (Section \ref{sec:basenis}).
To sort out these differences, we repeated the full set of numerical experiments after removing a key opacity unavailable in the previous studies--- the LMSH electron capture table.
The results using the IPA EC (Section \ref{sec:ipanis}) follow the general expectations of the previous studies.
The analysis of the difference made by the choice of EC opacity during collapse (Section \ref{sec:ecnis}) provides a cautionary warning about the coupled nature of the neutrino opacity contributions to collapse and the viability of explosions.

\subsubsection{NIS comparisons using LMSH EC table \label{sec:basenis}}

\begin{figure}
\epsscale{1.15}
\plotone{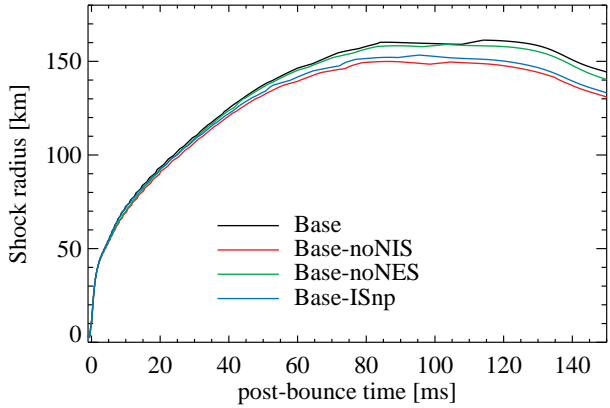}
\caption{Shock trajectories in km, versus time after bounce, for all models in Section \ref{sec:basenis}. The models plotted are \base\ (black; all opacities included), \basenis\ (red; without NIS opacities; no NES, no NPS, nucleon IS), \basenes\ (green; without NES), and \basereds\ (blue; nucleon IS). Shock position is computed by bisecting the pair of mass shells with the largest negative radial velocity gradient $-\partial v_r/\partial r$.  \label{fig:baseshock}  }
\end{figure}

\begin{figure}
\epsscale{1.15}
\plotone{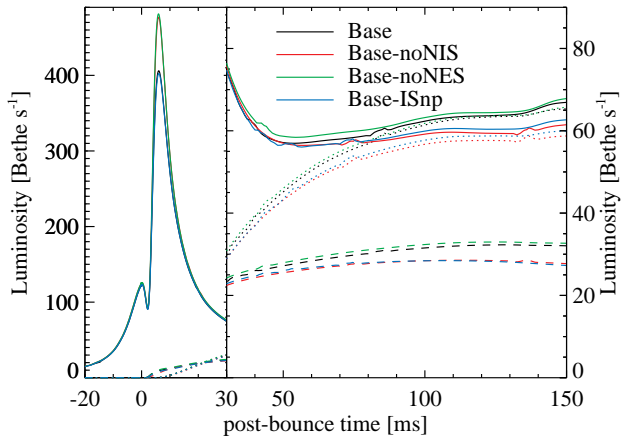}
\caption{Comoving-frame neutrino luminosities measured at 400~km for all models in Section \ref{sec:basenis}. Colors are as in Figure~\ref{fig:baseshock}. Electron neutrino, \nue, luminosities are represented by solid lines, \nuebar-luminosities by dotted lines, and \numt-luminosities by dashed lines. For the heavy lepton flavor neutrinos, $L_{\mu\tau} \equiv L_\mu = L_\tau$, is the luminosity of a single species and not a sum.  \numtbar-luminosities are indistinguishable from \numt-luminosities, and omitted from this figure. The luminosities are in \Bethes, where 1~Bethe = $10^{51}$~ergs.  \label{fig:baselumin}}
\end{figure}

\begin{figure}
\epsscale{1.15}
\plotone{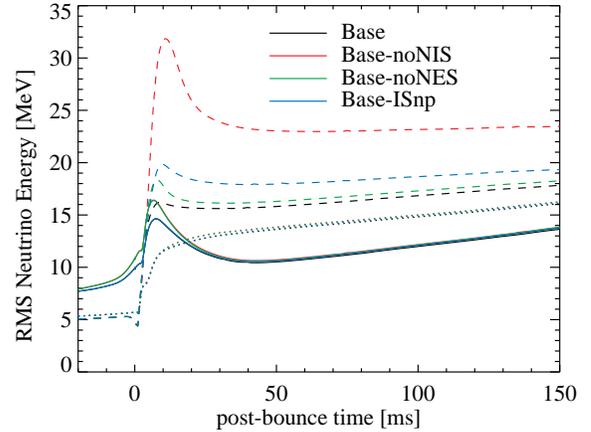}
\caption{Comoving-frame neutrino RMS energies, $\meanE{\nu} = (\int d\mu \, dE \, E^4 F/ \int d\mu \, dE \, E^2 F)^{1/2}$, measured at 400~km for all models in Section \ref{sec:basenis}. RMS energy is computed over number density, not number flux. Colors are as in Figure~\ref{fig:baseshock}. Line styles are as in Figure~\ref{fig:baselumin}. \label{fig:baseavgen} }
\end{figure}

For this set of tests we compare a model with our full opacity set (\base) to models without electron scattering (\basenes), without positron scattering (\basenps), replacing the nucleon scattering of \citet{RePrLa98} with the IS equivalent of \citet{Brue85} (\basereds), and to a model with all three of these changes (\basenis).
\emph{At bounce we find no noticeable differences among the models.} (See Figure~\ref{fig:bounce} and Table~\ref{tab:models}.)
As the shock reaches its largest extent (Figure~\ref{fig:baseshock}), two splittings are visible. 
(The model without positron scattering, \basenps, is omitted from the plots and discussion as it is indistinguishable from the \base\ model.)
The smaller splitting reaching 2--3 km by $\approx 80$~ms post-bounce  is created by omitting electron scattering.
The larger splitting, originating at $\approx 30$~ms post-bounce, is 10--15 km and is created by replacing the NIS nucleon scattering with the IS equivalent.
In both cases, the larger radius is obtained by the model with more NIS opacity.
The splitting in the shock radii is reflected in the luminosities for the same epoch (Figure~\ref{fig:baselumin}).
The choice of nucleon scattering opacity makes the largest difference, with the nucleon NIS-containing models (\base, \basenes) having larger luminosities for all neutrino species, $\approx$5--6~\Bethes\ for each species at 150 ms after core bounce, due to smaller total scattering opacity for the nucleon NIS of \citet{RePrLa98} than the nucleon IS of \cite{Brue85}, permitting easier escape of trapped \nue\nuebar\ from the core.
A much smaller difference in luminosities exists for the omission of NES. 
The larger \nue- and \nuebar-luminosities for models using the NIS nucleon scattering sustain larger shock radii through higher gain-region net heating rates.
For the smaller splitting in luminosity and shock radius due to removing NES, the correlation between luminosity and shock radius is weaker, being inverted in the case including nucleon NIS. We were unable to isolate a cause for this small difference.
For the  two models using nucleon IS, the model with NES (\basereds) has slightly higher luminosity and shock radius than the model without NES (\basenis); but, for the two models with nucleon NIS, the model with NES (\base) has {\it lower} luminosity and higher shock radius relative to the model without NES (\basenes).
\citet{MaJa09} have proposed an analytic proportionality for the shock radius, $R_{\rm sh} \propto ( L_\nu \langle E_\nu^2\rangle )^{4/9} R_{\nu \rm s}^{16/9}$, where $R_{\nu \rm s}$ is the neutrinosphere radius. If we consider only the effect of \nue\nuebar-luminosity we would find a relation closer to $R_{\rm sh} \propto  L_\nu$ for the scattering models showing larger differences in luminosities in this section and Section~\ref{sec:ipanis}, indicating that the increase in neutrinosphere radii from lower opacities (for the more luminous models) are also playing a role in the shock radius.
There are no detectable differences among these models in the \meanE{\nue}\ and \meanE{\nuebar}\ (Figure~\ref{fig:baseavgen}) during the accretion epoch.
During the accretion phase, the various neutrinospheres and the entire gain region are within a gas composed primarily of electrons and free protons and neutrons, and the nucleon scattering opacity improvements are reflected in the radiation and hydrodynamic quantities.
In addition to permitting non-isoenergetic scattering, the improved nucleon opacities account for nucleon phase blocking, recoil, etc., which also make modest alterations to the total scattering opacity.

During breakout we see a different behavior, with NES being the more important NIS opacity.
The breakout burst $L_{\nue}$ is approximately 480~\Bethes\ for models without NES (\basenes, \basenis) and approximately 405~\Bethes\ for models with NES (\base, \basereds).
The same grouping applies for \meanE{\nue}, which is up to 2~\mev\ higher during breakout and up to 1~\mev\ higher before bounce for models without NES.
The breakout burst represents the passage of the shock through the neutrinospheres, and the material above the shock is primarily composed of heavy nuclei.
Therefore, scattering on free nucleons is of less importance to the neutrino spectrum during breakout.
The electrons above the shock  downscatter the energies of the escaping neutrinos and lower the \meanE{\nue}\ and, therefore, the total luminosity.

The most dramatic difference seen in these models is in the quantity \meanE{\numt}\ (Figure~\ref{fig:baseavgen}, dashed lines), with significant differences across all the models. 
At 150~ms post-bounce the increase in \meanE{\numt} relative to the \base\ model is approximately 0.5 and 1.5~\mev, respectively, for models \basenes\ and \basereds, while the increase is nearly 7~\mev\ for model \basenis, with all NIS scatterings removed.
This demonstrates the non-linear nature of the opacity changes.
Energy lost by \numt\ via scattering thermalizes the spectra of \numt\ reducing \meanE{\numt} and serves as an important source of heating between the \numt- and \nue-neutrinospheres. (See Section~\ref{sec:pair} for further discussion.)
In model \basenis\ there are no scattering processes remaining to thermalize the spectrum of \numt\ after emission.
NIS from either nucleons or electrons is enough to significantly lower \meanE{\numt}, with both required for the full effect.
The lack of variation in \meanE{\nue\nuebar} after breakout for these models demonstrates that absorption followed by emission plays the role of an ``effective downscattering'' in thermalizing the neutrino spectra.

\subsubsection{NIS comparisons using IPA EC \label{sec:ipanis}}

\begin{figure}
\epsscale{1.15}
\plotone{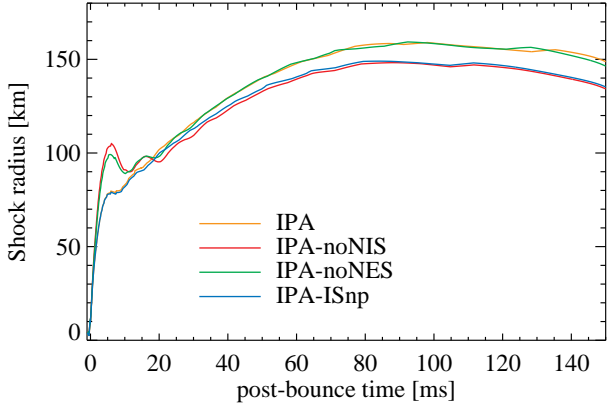}
\caption{Shock trajectories in km, versus time after bounce, for all models in Section \ref{sec:ipanis}. Models are \ipa\ (orange; all NIS opacities), \ipanis\ (red; without NIS opacities; no NES, no NPS, nucleon IS), \ipanes\ (green; without NES), and \ipareds\ (blue; nucleon IS).  \label{fig:ipashock}}
\end{figure}

\begin{figure}
\epsscale{1.15}
\plotone{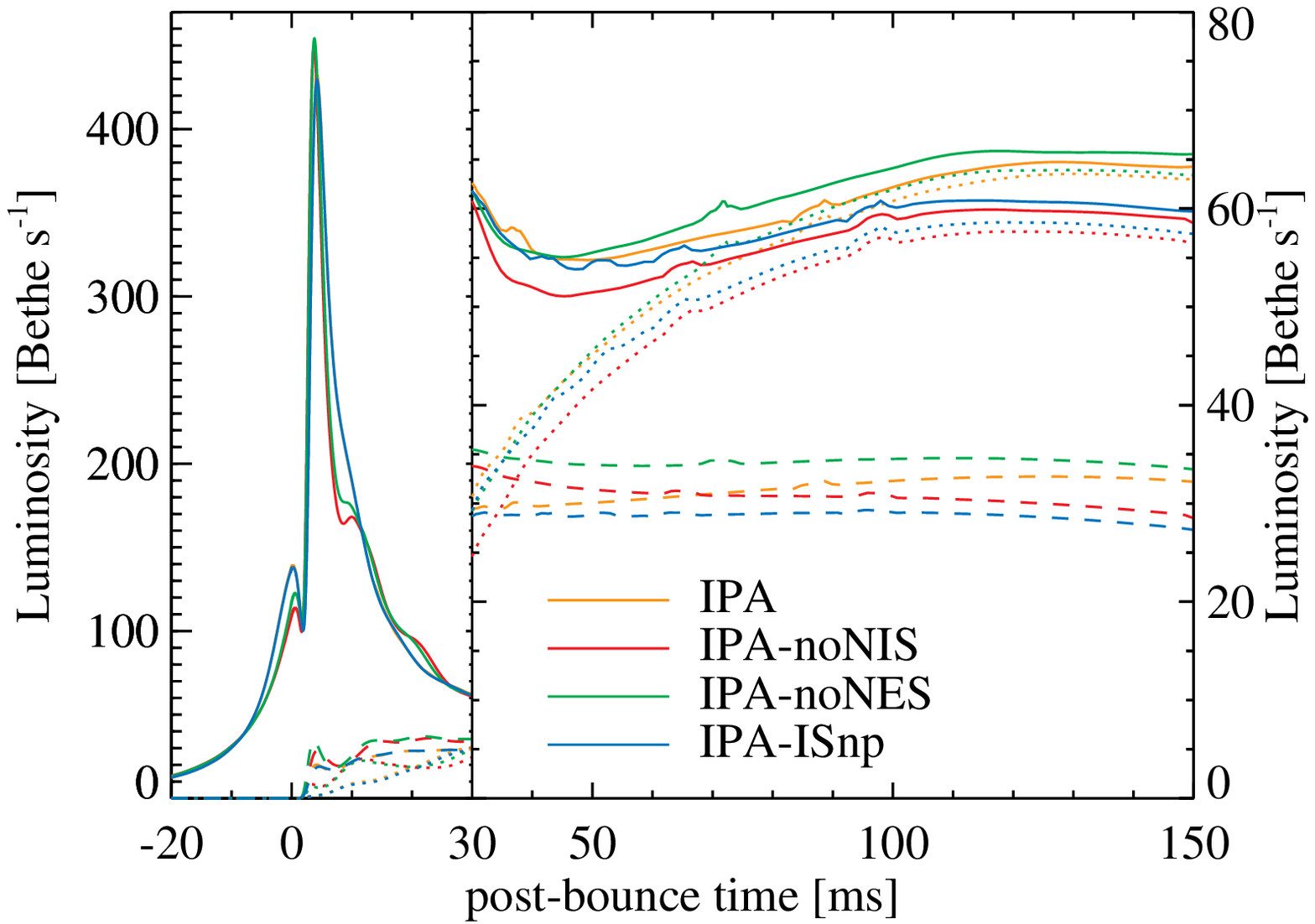}
\caption{Comoving-frame neutrino luminosities measured at 400~km for all models in Section \ref{sec:ipanis}. Colors are as in Figure~\ref{fig:ipashock}.Line styles are as in Figure~\ref{fig:baselumin}. The luminosities are in \Bethes, where 1~Bethe = $10^{51}$~ergs. \label{fig:ipalumin}}
\end{figure}

\begin{figure}
\epsscale{1.15}
\plotone{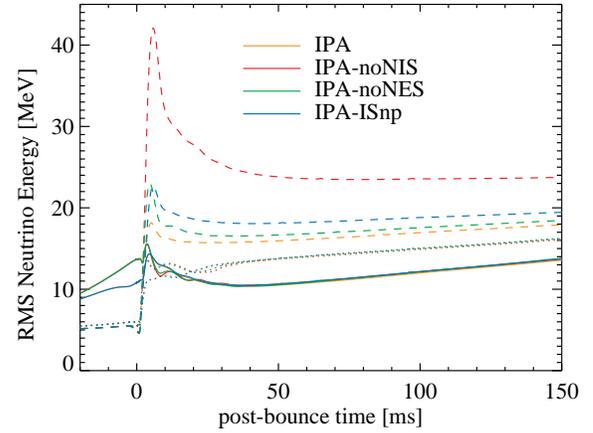}
\caption{Comoving-frame neutrino RMS energies measured at 400~km for all models in Section \ref{sec:ipanis}. Colors are as in Figure~\ref{fig:ipashock}. Line styles are as in Figure~\ref{fig:baselumin}.  \label{fig:ipaavgen} }
\end{figure}

The lack of effects at bounce from reduced NIS opacities stands in contrast to the results of \citet{MeBr93c} who used an earlier version of our code with the same NES opacity.
Therefore, we repeated our NIS subtraction experiments with the LMSH EC table replaced by the IPA EC of \citet{Brue85} that \citet{MeBr93c} used.
The replacement of the LMSH EC table with the IPA EC has the largest effect on the bounce configuration of any pair or emission/absorption opacity alternatives. (See Figure~\ref{fig:bounce}. Those differences will be discussed individually in Sections~\ref{sec:abem} and~\ref{sec:pair}.)
At core bounce (Figure~\ref{fig:bounce}), the primary differences among these models are attributed to the inclusion or omission of NES.
The homologous core mass, \Mshock, shifts from 0.544~\msun for model  \ipa\  with all NIS opacities, to 0.608~\msun\ for model \ipanes\  with NES removed.
There is a slightly larger shift in \Mshock\ to 0.618~\msun\ for model  \ipanis\ without the remaining NIS opacities.
(The model without NPS, \ipanps, is indistinguishable from the  model \ipa\ throughout and is omitted from the plots and discussion.)
Correlated to the increase in \Mshock\ is  $Y_L$, which  increases from 0.333 for \ipa\ to 0.365 for \ipanes.
There is a corresponding increase in $Y_e$ from 0.284 to 0.306 and in net neutrino number, $\Ynu = Y_L - Y_e$, from 0.0488 to 0.0591, consistent with the effects of NES described by \citet{MeBr93c}, where scattering by NES moved neutrinos to lower energies with lower opacities and an easier path to escape.
In the models without NES, neutrinos are not scattered to lower energies by electrons, and are not aided in their escape.
During collapse, the core consists primarily of heavy nuclei and electrons, therefore NIS on free nucleons has little effect, as we can see from these models.
The model with nucleon IS and including NES, \ipareds, is indistinguishable from model \ipa\ at bounce.
The models with higher core $Y_e$ have a correspondingly higher core density, pressure, and post-shock temperature.

During breakout, the inclusion or omission of NES remains the primary difference among the models.
The shocks (Figure~\ref{fig:ipashock}) for these models with the IPA EC on nuclei launch more vigorously and from shallower depths in the gravitational well  than the shocks for the LMSH EC models (Figure~\ref{fig:baseshock}).
The vigor of these launched shocks creates a slight oscillation in shock radius for the models with NES (\ipa, \ipareds) and a prominent shock ``ringing'' in the models without NES (\ipanes, \ipanis), also seen by \citet{ThBuPi03} for a model without NES.
The breakout luminosity for \nue\ peaks somewhat higher at $L_{\nue} \approx 450$~\Bethes\ for the models without NES and $\approx 430$~\Bethes\ for the models that include NES. (See Table~\ref{tab:models}.)
The breakout burst luminosity in \nue\ also drops faster for the models without NES and exhibits oscillations in all luminosities and \meanE{\nu}\ (Figure~\ref{fig:ipaavgen}).
The RMS energies for \nue\ are also higher in the pre-bounce phase for the models without NES.
During breakout the shock passes through the various \nue-neutrinospheres, and in the case of the models without NES, the shock oscillates through the neutrinospheres.

At 20~ms after bounce the shock trajectories cross (Figure~\ref{fig:ipashock}) and by 40~ms after bounce the primary difference between models is the use of NIS or IS nucleon opacities.
Like for the models with the LMSH table, the shock radius in the later epochs is $\approx$10--12~km larger for models using the nucleon NIS of \citet{RePrLa98} (\ipa, \ipanes) than those using the nucleon IS of \citet{Brue85} (\ipareds, \ipanis).
Within these pairs of models and during this epoch, the difference caused by NES is even smaller for the IPA~EC models than the LMSH EC models.
This difference in shock radii is also reflected in the luminosities (Figure~\ref{fig:ipalumin}), with larger luminosities for models with nucleon NIS  than for those with nucleon IS.
Lower scattering opacity for nucleon NIS again enhances escape of trapped \nue\nuebar, increasing their luminosities, the heating from absorption, and the shock radius.
The smaller shifts in luminosity within these nucleon scattering pairs due to their differences in the inclusion or omission of NES are not reflected in the shock radii.

The late-time behavior of \meanE{\nu}\ (Figure~\ref{fig:ipaavgen}) for these IPA EC models is similar to the LMSH EC models.
After $\approx 40$~ms, differences in \meanE{\nue}\ and \meanE{\nuebar}\ are undetectable for these models.
The peak \meanE{\numt} (dashed lines) during breakout changes from 18~\mev\ for model \ipa\ with all NIS opacities included, to approximately 23~\mev\ for the two models with one NIS opacity missing (\ipanes, \ipareds), to 42~\mev\ for model \ipanis\ with all NIS opacities omitted.
For the \ipanis\ model, there are no energy-exchanging scatterings to alter the spectrum, and what we see is the emission spectra of all emitted neutrinos integrated over the semi-transparent pair-emitting region.
Inclusion of either NES or nucleon NIS is enough to push the observed \numt-spectrum most of the way toward the values in model \ipa\ with all of the NIS opacities.
A similar effect is also seen at 150~ms after bounce when the models with one missing NIS opacity (\ipanes, \ipareds) have \meanE{\numt}\ that is $< 2$~\mev\ larger than for the \ipa\ model, while for the model without any NIS opacities (\ipanis), \meanE{\numt}\ is 6~\mev\ larger than for model \ipa, demonstrating the non-linear interplay of the NIS opacities on the thermalization of the \numt\numtbar\ spectra.

\subsubsection{Interaction between NIS and EC during collapse \label{sec:ecnis}}

\begin{figure}
\epsscale{1.2}
\plotone{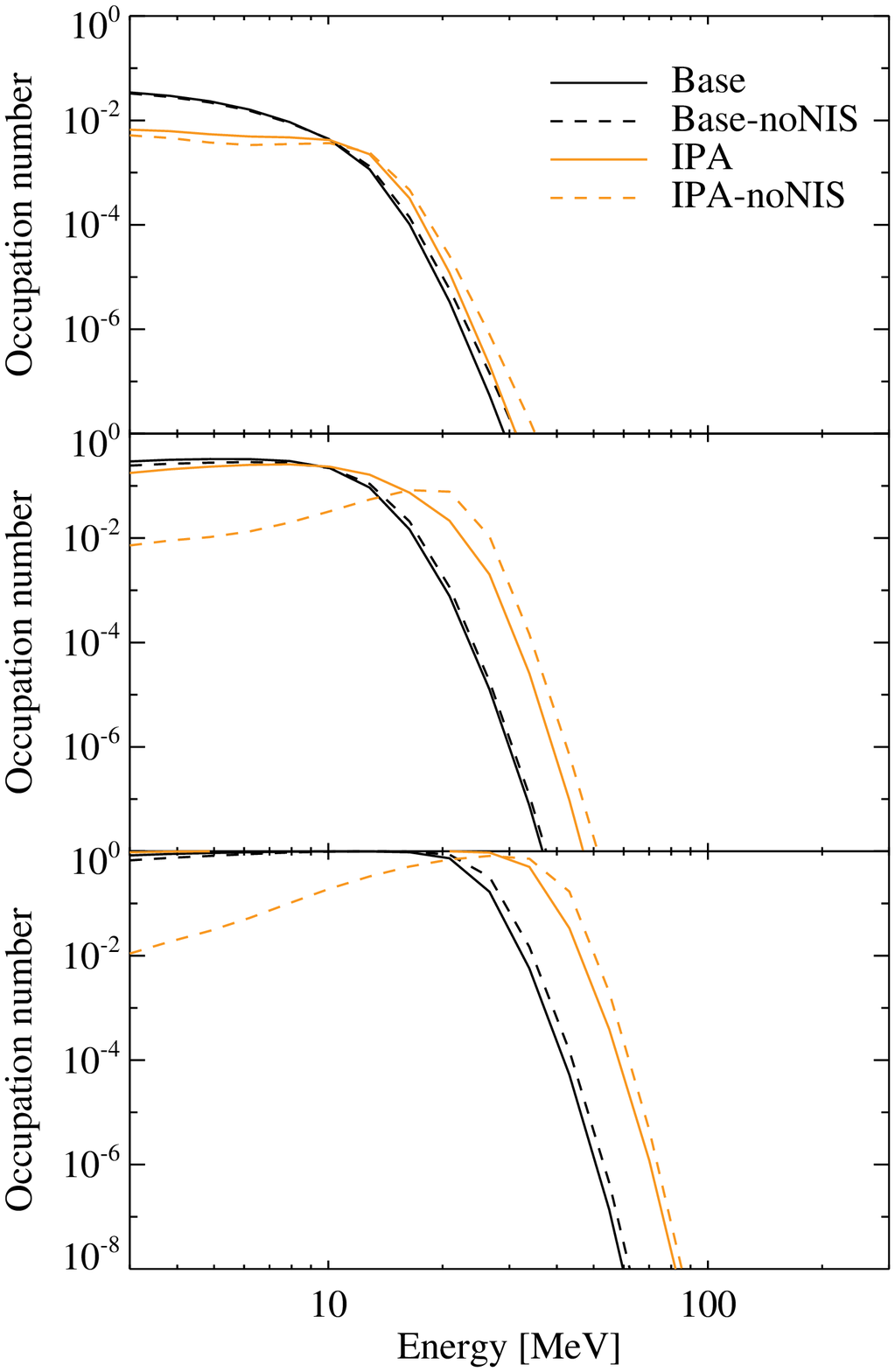}
\caption{Spectral neutrino occupation number (zeroth angular moment of the invariant neutrino distribution function) versus comoving-frame neutrino energy  at mass shell $M=0.25\,\msun$ for the models discussed in Section \ref{sec:ecnis}. The models are \base\ (black, solid), \basenis\ (black, dashed), \ipa\ (orange, solid), and \ipanis\ (orange, dashed). The three panels are for the models at central densities, $\rho_c$, of \den{1}{11}\ (upper panel), \den{1}{12}\ (middle panel), and \den{1}{13}\ (lower panel).  \label{fig:spectrum} }
\end{figure}

To illustrate the interplay between nuclear electron capture and non-isoenergetic scattering during collapse, we plot the occupation number (where values of 1 represent a completely occupied phase-space) of \nue\ as a function of energy,  the zeroth moment of the distribution function, for three pre-bounce epochs.
In  Figure~\ref{fig:spectrum} we compare the spectra of models with the LMSH EC (\base, \basenis; black lines) to models using the IPA EC (\ipa, \ipanis; orange lines), for both cases: including the full set of NIS opacities (\base, \ipa; solid lines) and omitting all NIS opacities (\basenis, \ipanis; dashed lines).
During collapse the central density, $\rho_c$, serves as a useful ``clock'' for comparing different models, and the spectra are plotted for $\rho_c = 10^{11}, 10^{12}$, and $10^{13}\, \gcc$ (top, middle, and bottom panels, respectively).
At $\rho_c = 10^{11}\,\gcc$ (upper panel) the spectra are relatively similar, though the IPA EC models have fewer neutrinos at low energies and slightly more neutrinos in the high-energy tail.
At $\rho_c = 10^{12}\,\gcc$ (center panel) there is a clear separation between the models.
The overall phase-space occupation has increased for all models, though for the lowest energy neutrinos in model \ipanis, it has only slightly increased.
The high-energy tail for the IPA EC models (orange lines) is shifted  $\approx$~40--50\% higher in energy relative to the LMSH EC models (black lines), with the model lacking NIS, \ipanis, having the largest shift.
At energies less than $\approx 15\,\mev$, model \ipanis\ has fewer neutrinos than the other models, approximately 30 times fewer for the lowest-energy neutrinos.
Unlike the other models, which have fairly flat neutrino spectra up to the high-energy roll-off, model \ipanis\ has a peak at 20~\mev.
At $\rho_c = 10^{13}\,\gcc$ (lower panel) the trend continues with model \ipanis\ as the outlier.
The high-energy tail of the \nue-spectrum for IPA EC models (orange lines) is again shifted  $\approx$~40--50\% in energy relative to the LMSH EC models (black lines) with a smaller shift for models without NIS (dashed lines) relative to those with NIS (solid lines).
The neutrino phase-space is (nearly) completely filled for $E < 20\,\mev$ for the LMSH EC models (black lines) and for $E<30\,\mev$ for model \ipa, while model \ipanis\ reaches a peak  at 30~\mev\ and decreases for lower energies.
The spectra for the IPA EC models (Figure~\ref{fig:spectrum}, orange lines) are consistent with previously reported collapse-phase spectra \citep{Brue85,SmCevdH96}.

In  the independent particle approximation, nuclear EC is completely shut-off when the mean neutron number $N > 40$, which occurs at densities above $\approx \den{2}{10}$, constituting much of the core collapse.
Therefore, in the IPA EC models (\ipa, \ipanis) all of the \nue\ emission in regions with density exceeding $\approx \den{2}{10}$ arises from electron capture on free protons, which are relatively rare, and results in a slower overall rate of \nue\ emission and core deleptonization. (Pair emission is correspondingly low during collapse as we shall discuss in Section~\ref{sec:pair}.)
The LMSH EC implementation emits neutrinos with a lower mean energy than the capture on protons, which dominates the total capture rate in the IPA models at densities above $\approx \den{2}{10}$, and fills the low energy spectra which is underpopulated by EC on protons \citep{LaMaSa03}, which we can see by comparing the spectral evolution of the models without NIS opacities for the LMSH EC (\basenis) and IPA EC (\ipanis).
With IPA EC, the low-energy  spectrum can only be filled by energy downscattering by NIS opacities, which efficiently fill the low-energy spectrum for model \ipa.
Using the LMSH EC fills the low-energy \nue\ spectrum directly without  energy downscattering (model \basenis), so we do not see an enhancement in neutrino escape through the less opaque neutrino ``window'' at lower energies when NIS opacities are included, like we do for the IPA EC models.

Scattering on nuclei is the dominant contributor to total opacity during collapse and identical in all of our models.
Once the low-energy phase space is filled, deleptonization is controlled by the total opacity, and thus there are no discernible differences in the deleptonization among the LMSH EC models presented in Section~\ref{sec:basenis}.
For the IPA EC models presented in Section~\ref{sec:ipanis}, changes to the filling of the low-energy phase space, do affect the net escape after trapping begins.
The differences between the models with the full set of NIS opacities (\base, \ipa) are a consequence of differences in EC before trapping occurs \citep{HiMeMe03}.

After considering the effects of compression due to variation in central density on the temperature and the change in entropy profiles due changes in \Mshock\ among the models plotted in Figure~\ref{fig:bounce}, there is likely an additional small increase in temperature and entropy inside the shock for models that include IPA EC and NES (orange lines). This small thermal increase is likely due to energy gained by the fluid from the scattering of \nue\ by electrons as can be seen in difference between the model \ipa\ and model \ipanis\ spectra in Figure~\ref{fig:spectrum}. The lack of similar effects on the entropy and temperature by changes in scattering on models using the LMSH EC, suggests (as does the lack of impact on the spectra during collapse) that the NIS has very little impact on  \nue\ during collapse when utilizing the more complete LMSH EC implementation.

\subsection{Emission and absorption comparisons \label{sec:abem}}

\begin{figure}
\epsscale{1.15}
\plotone{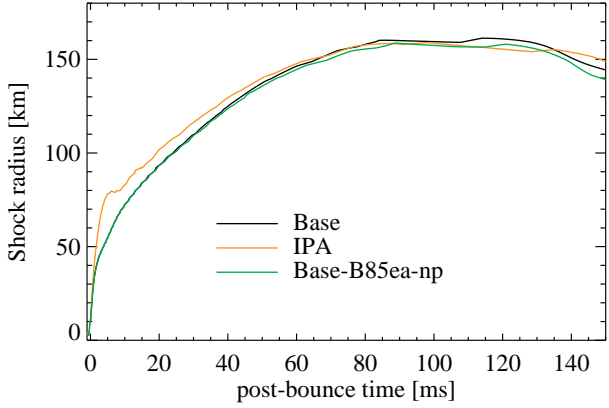}
\caption{Shock trajectories in km, versus time after bounce, for all models in Section \ref{sec:abem}. Models are \base\ (black; all opacities), \ipa\ (orange; IPA EC), and \baserea\ (green; B85 EC on free nucleons). \label{fig:abemshock}}
\end{figure}

To test the two sources of \nue\ (\nuebar) emission via electron (positron) capture and the inverse neutrino absorption in the modern opacity set of model \base, we replaced them  with their \cite{Brue85} equivalents for EC on nuclei (model \ipa) and EC on free nucleons (model \baserea).
At bounce (Figure~\ref{fig:bounce}), the change in nucleon EC produces no discernible difference between models \base\ and \baserea, but changing the nuclear EC produces the largest differences at bounce of any single opacity replacement tested in this paper.
As noted by \citet{HiMeMe03}, the IPA EC (model \ipa) results in less deleptonization, as the IPA turns off completely where the heavy nuclei have neutron numbers $N\ge 40$, of the inner core relative to using the LMSH EC (model \base) leading to higher $Y_e$, $Y_L$, density, pressure, and temperature inside the bounce shock.
The IPA EC also results in higher deleptonization and stronger collapse \emph{outside} the bounce shock, which can be seen in the lower $Y_e$, higher density, and higher infall velocities  for model \ipa\ outside the bounce shock (Figure~\ref{fig:bounce}) where the stronger IPA for low density does not shut off. This will increase the ram pressure as the shock passes through the Fe-core.
The shock forms at 0.554~\msun\ in model \ipa\ and 0.430~\msun\ in model \base, and the shallower shock launch results in a more vigorous launch of the shock for model \ipa\ (Figure~\ref{fig:abemshock}).
Approximately 70~ms after bounce the shock trajectories of the two nuclear EC models cross  \citep[as reported by][]{HiMeMe03} due to higher ram pressure from higher density and infall velocities in the outer core of model \ipa\ induced by stronger deleptonization.
The shock trajectories of models \base\ and \ipa\ cross again late in our simulations, with the net effect that the shock position of model \ipa\ is flatter during the epoch when we should expect multidimensional effects to become important.
In the model using the B85 nucleon EC (\baserea), the shock radius trails that in the \base\ model slightly, with the deficit reaching 5~km by the end of our simulations at 150~ms post-bounce.
This deficit in shock radius for model \baserea\ is reflected in an $\approx 2\, \Bethes$ lower \nue- and \nuebar-luminosity during the accretion phase relative to model \base, with a small (15~\Bethes) decrease in the breakout \nue-luminosity.
The shallower bounce shock of model \ipa\ also results in a breakout \nue-luminosity burst that peaks sooner and higher (by $\approx 25\, \Bethes$) than the LMSH EC model (\base), and then drops sooner creating a narrower breakout peak.
Model \ipa\ shows a higher \numt-luminosity from bounce to approximately 80~ms post-bounce.
With no discernible differences in \meanE{\numt}\ outside the narrow peak during breakout (Figure~\ref{fig:abemavgen}), the higher luminosity of model \baserea\ implies it is emitting more \numt\numtbar-pairs than the other models.
The \meanE{\nu}\ for all models is essentially identical for each neutrino species during the accretion epoch, with sharper and narrower peaks for model \ipa\ during breakout, owing to the shallower and rapidly moving shock.
Before bounce, model \ipa\ shows a 1--2~\mev\ higher \meanE{\nue}, reflecting the shift of the high-energy tail seen in Figure~\ref{fig:spectrum} and discussed in Section~\ref{sec:ecnis}.

\begin{figure}
\epsscale{1.15}
\plotone{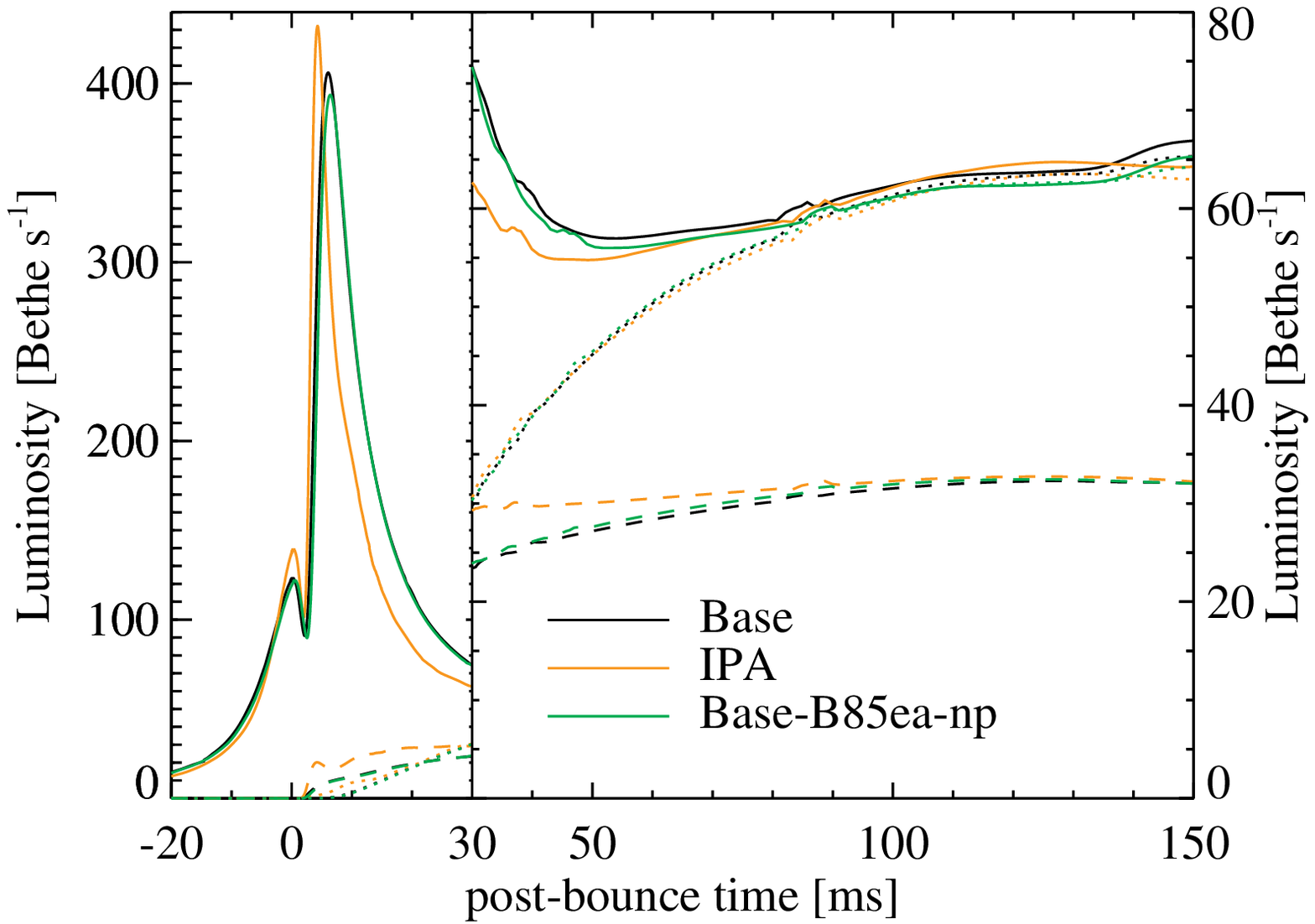}
\caption{Comoving-frame neutrino luminosities measured at 400~km for all models in Section \ref{sec:abem}. Colors are as in Figure~\ref{fig:abemshock}. Line styles are as in Figure~\ref{fig:baselumin}. The luminosities are in \Bethes, where 1~Bethe = $10^{51}$~ergs. \label{fig:abemlumin}}
\end{figure}

\begin{figure}
\epsscale{1.15}
\plotone{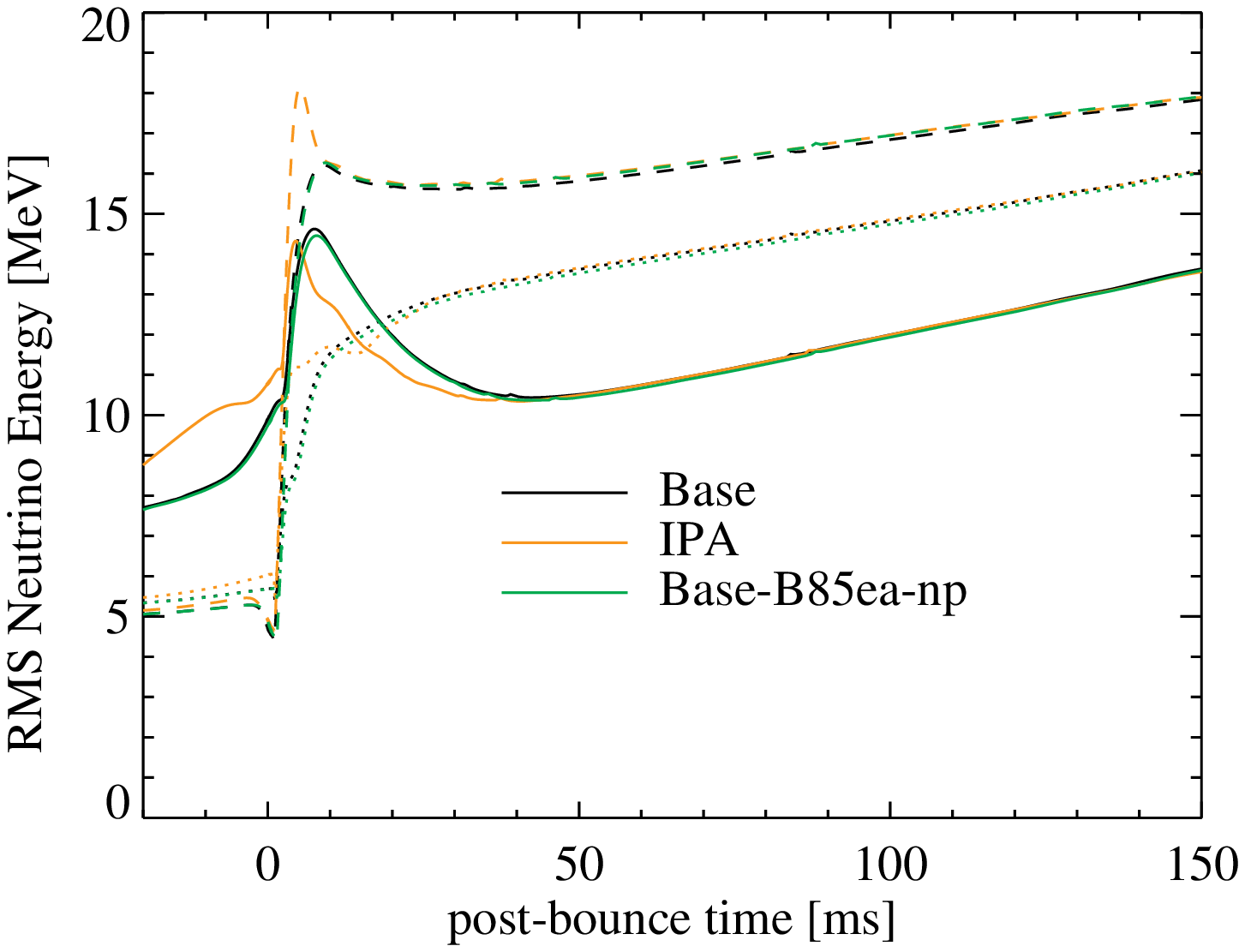}
\caption{Comoving-frame neutrino RMS energies measured at 400~km for all models in Section \ref{sec:abem}.  Colors are as in Figure~\ref{fig:abemshock}. Line styles are as in Figure~\ref{fig:baselumin}. \label{fig:abemavgen} }
\end{figure}

\subsection{Pair opacity comparisons\label{sec:pair}}

\begin{figure}
\epsscale{1.15}
\plotone{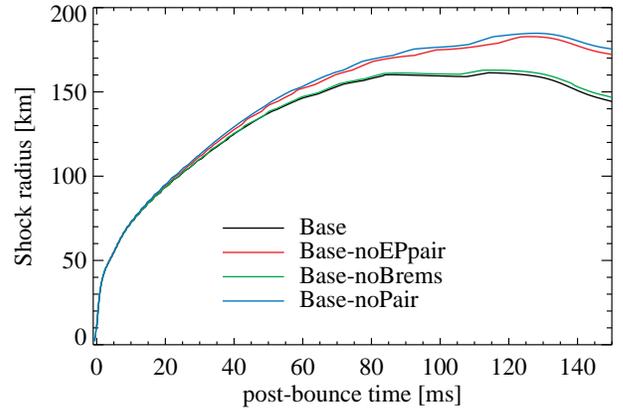}
\caption{Shock trajectories in km, versus time after bounce, for the pair opacity test models in Section \ref{sec:pair}. The models are \base\ (black; all opacities), \basepair\ (red; without $e^+e^-$-annihilation),  \basebrem\ (green; without Bremsstrahlung), and \basenopair\ (blue; without pair opacities). \label{fig:pairshock}}
\end{figure}

To test the effect of pair opacities on the supernova models, we compare the \base\ model to models omitting $e^+e^-$ annihilation (\basepair),  bremsstrahlung (\basebrem), and both (\basenopair) pair sources.
Without any pair sources model \basenopair\ also lacks  \numt\numtbar.
At bounce there are no distinguishable differences among these models, which is consistent with the lack of thermal $e^+e^-$ pairs to annihilate and free nucleons for bremsstrahlung.
The shock trajectories for the models with missing pair opacities begin to deviate from model \base\ approximately 40~ms after bounce (Figure~\ref{fig:pairshock}), with the model \basepair\ shock extending to 183~km, the model \basebrem\ shock extending to 163~km, and the model \basenopair\  shock extending to 185~km.
These are the only  models in this paper that exceed the maximum shock extent of the \base\ model at 161~km. (See Table~\ref{tab:models}.)
Comparison of these models shows that  $e^+e^-$ annihilation is  much more important to the shock propagation than bremsstrahlung.

\begin{figure}
\epsscale{1.15}
\plotone{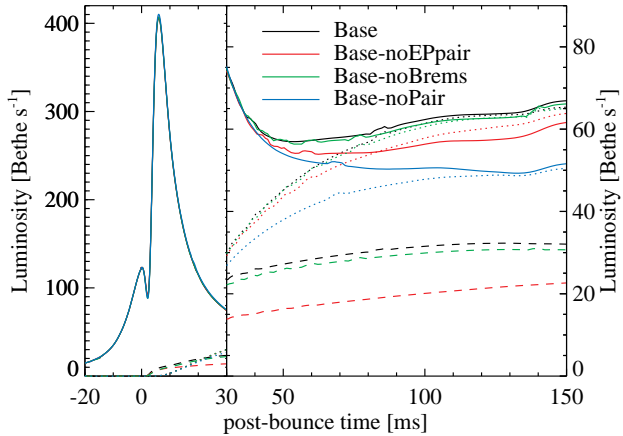}
\caption{Comoving-frame neutrino luminosities measured at 400~km for all models in Section \ref{sec:pair}. Colors are as in Figure~\ref{fig:pairshock}. Line styles are as in Figure~\ref{fig:baselumin}. The luminosities are in \Bethes, where 1~Bethe = $10^{51}$~ergs. The dashed line for model \basenopair\ $L_{\numt}$ is omitted as this model does not include \numt\numtbar. \label{fig:pairlumin}}
\end{figure}

\begin{figure}
\epsscale{1.15}
\plotone{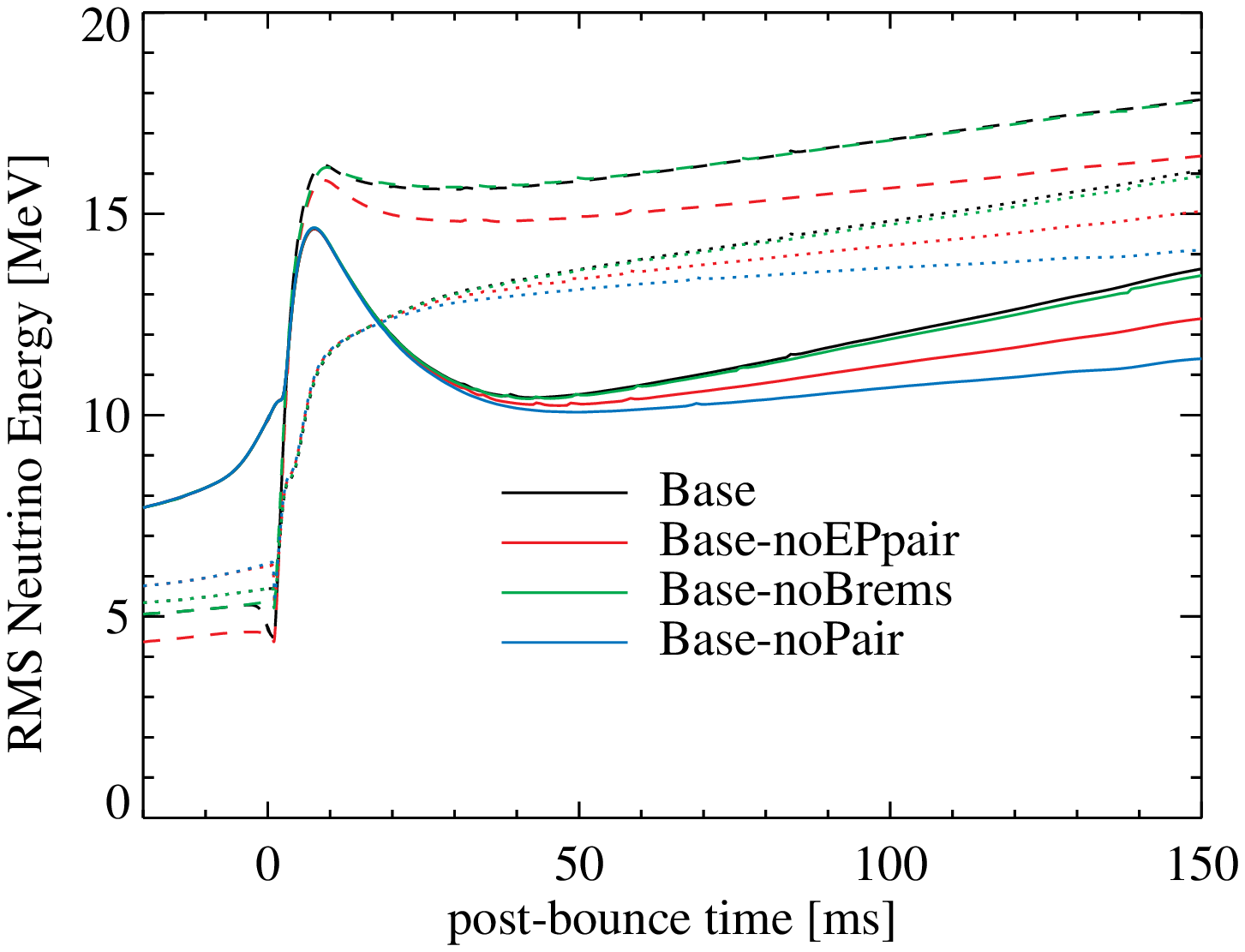}
\caption{Comoving-frame neutrino RMS energies measured at 400~km for all models in Section \ref{sec:pair}. Colors are as in Figure~\ref{fig:pairshock}. Line styles are as in Figure~\ref{fig:baselumin}. \label{fig:pairavgen} }
\end{figure}

Unlike the previous comparisons, where accretion-phase shock differences are correlated with \nue- and \nuebar-luminosities, the luminosities (Figure~\ref{fig:pairlumin}) for these models are \emph{anti-correlated} with the shock radii.
The differences in the \meanE{\nu}\ for all neutrino species (Figure~\ref{fig:pairavgen}) are also anti-correlated, with the shock with smaller differences relative to model \base\ for model \basebrem\ and larger differences for model \basepair.
In model \basenopair\ the only source of \nue\ is electron capture on protons and of \nuebar\ is positron capture on neutrons.
As in the case of  the EC--NIS interplay (Section~\ref{sec:ecnis}), the non-linearity of pair opacity effects on the emission of \nue\nuebar\ is clear.
For both luminosity and RMS energy the removal of bremsstrahlung had only a minor impact when  $e^+e^-$-annihilation was present.
However, when $e^+e^-$-annihilation was absent, the removal of bremsstrahlung had a significant impact on both.

\begin{figure}
\epsscale{1.15}
\plotone{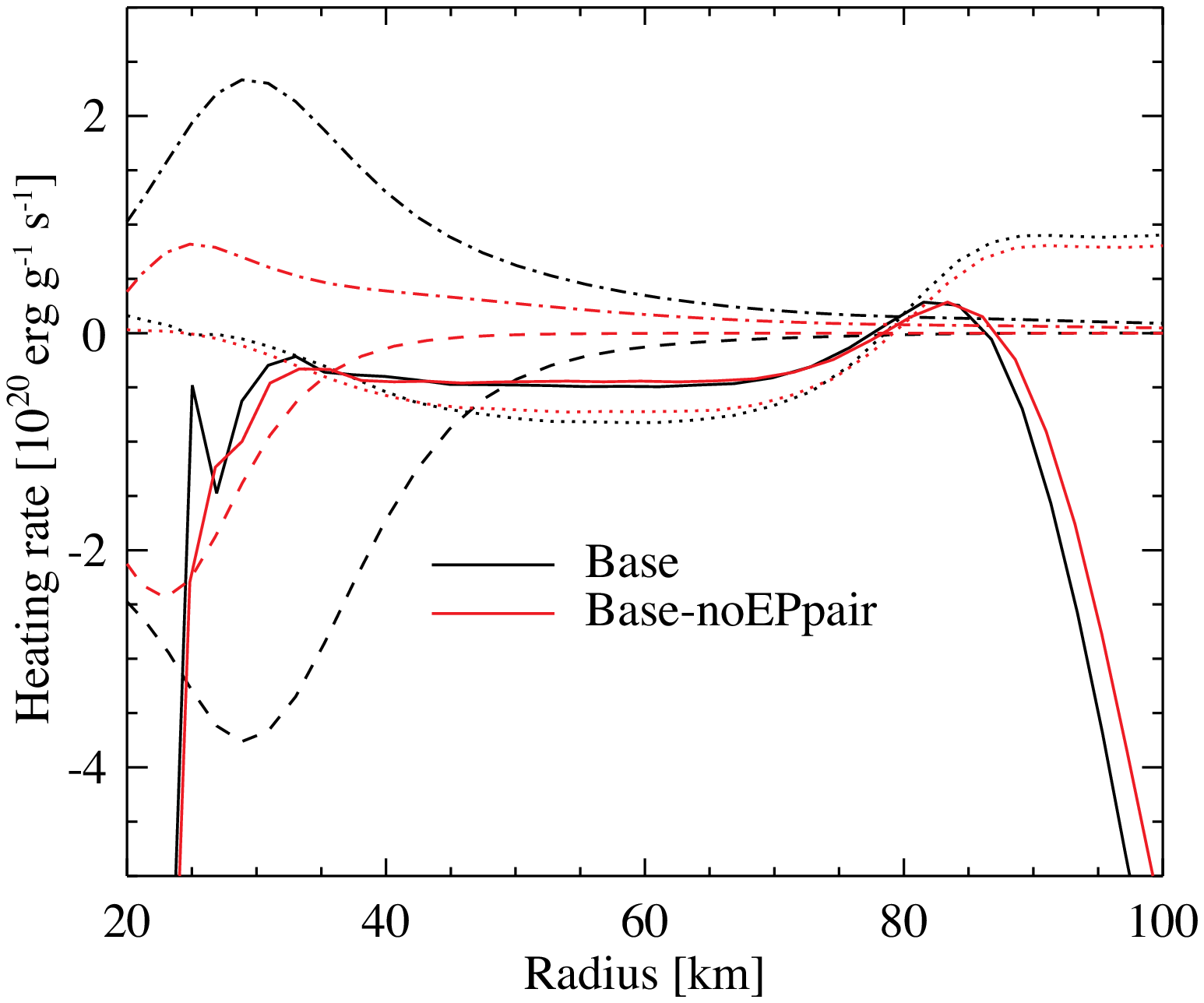}
\caption{Net heating rates in $\mbox{ergs g}^{-1}\,{\rm s}^{-1}$ for models \base\ (black) and \basepair\ (red) at 30~ms after bounce. Line styles indicate net heating by: emission/absorption by \nue\ (solid) and \nuebar\ (dotted), by \numt\numtbar\ pair sources (dashed) and by NIS of \numt\numtbar\ (dash-dotted). Net pair  heating for \nue\nuebar\ is approximately one order of magnitude or more smaller than the \numt\numtbar\ counterpart and is omitted from this plot for clarity. \label{fig:pairheat} }
\end{figure}

Previous studies \citep{ThBuHo00,ThBuPi03,BuYoPi00,KeRaJa03,BuJaKe03,BuRaJa06} have stressed the importance of bremsstrahlung as a neutrino-pair source.
\citet{BuRaJa06} plotted (see their Figures~21--22)  each  opacity for all neutrino species at two energies for two post-bounce epochs within our study range and found that bremsstrahlung was the dominant pair-source \emph{in the proto-neutron star}, but $e^+e^-$ annihilation became dominant starting somewhere outside the neutrinosphere in each case.
The \numt\numtbar\ emission rates for our models are consistent with that finding, being higher in the inner core for the bremsstrahlung-including models (\base, \basepair) than the model without bremsstrahlung (\basebrem).
Stationary transport studies \citep{ThBuHo00,BuYoPi00,KeRaJa03} have found bremsstrahlung to be more important than  $e^+e^-$ annihilation to the emitted flux and thermalization rate of \numt\numtbar. Stationary transport studies drive toward a global equilibrium that is never reached during the dynamic shock revival phase and can overemphasize the region below the neutrinospheres where bremsstrahlung is dominant.
A more relevant comparison is with the dynamic models of \citet{ThBuPi03} computed with and without bremsstrahlung like our models, though with different progenitor, base opacities, code, etc.
\citet{ThBuPi03} find that the total luminosity for all heavy lepton neutrinos, $L_x = 2 L_{\mu\tau} + 2 L_{\bar \mu\bar\tau}$, increases by 8--10~\Bethes\ starting from about 30~ms after bounce onward.
Our models are consistent with this change, showing an increase of  2~\Bethes\ for \emph{each} of the four heavy lepton (anti)neutrino species during the same epoch. The proportionate increase due to adding bremsstrahlung does appear larger for the \citet{ThBuPi03} models as their model without bremsstrahlung has roughly half the neutrino luminosity for each species as our \basebrem\ model. Much of the difference can be attributed to our use of nucleon NIS and GR with the rest due to other differences between the opacities, progenitors, and codes.
At later epochs, beyond the effective use of spherical models, the diffusion of core \numt\numtbar\ to the neutrinospheres may enhance the relative contribution of bremsstrahlung to the luminosity.

Removing pair sources reduces cooling by \numt\numtbar\ emission during the accretion phase.
Figure~\ref{fig:pairheat} shows the net heating rate  for models \base\ and \basepair\ at 30~ms after bounce, with the net cooling for \numt\numtbar\ emission shown as dashed lines. For the \base\ model, the cooling by \numt\numtbar\ emission dominates the radiative heat budget between 25 and 40~km, with about half of the cooling returned back to the local thermal pool by NIS on \numt\numtbar.
(In both of these models the full set of NIS opacities are available for thermalization during scattering.)
When $e^+e^-$-annihilation is removed (\basepair, red lines), both the pair cooling and related NIS heating are reduced and shifted deeper into the core.
The lower effective cooling by \numt\numtbar-pairs in model \basepair\ results in more thermal energy being available to support the shock and a larger shock radius.
Emission of \nue\nuebar-pairs is suppressed by the largely filled \nue\ phase space.

\section{Conclusions}

In this paper we have systematically examined the effects of each of the updated opacities in our modernized opacity set over the initial 150~ms post-bounce, spherically symmetric phase of core-collapse supernovae.
To summarize our primary findings:
(1) During collapse,  electron capture on heavy nuclei dominates the emission of neutrinos and the deleptonization of the core.
If modern EC rates are used with a detailed NSE composition, as in LMSH or \citet{JuLaHi10}, the direct emission of low-energy neutrinos obviates the need for NES to fill the low-energy portion of the neutrino spectrum.
(2) Omitting NES results in an $\approx 15$\% increase in the breakout-burst \nue-luminosity.
(3) Changes in shock formation due to deleptonization are the primary opacity-driven source of variation in early shock evolution.
(4) During the accretion phase, nucleon NIS enhances the \nue\nuebar-luminosities, net heating, and drives the shock further out, enhancing the potential of  shock  revival by multidimensional effects.
(5) Cooling by \numt\numtbar-pair emission from $e^+e^-$-annihilation removes energy from the system that could otherwise be used to revive the shock.
(6) All of the NIS opacities (except scattering on positrons) and $e^+e^-$-annihilation affect the neutrino luminosities and/or \meanE{\nu}\ during accretion.
(7) Positron scattering shows no impact on the outcome or observables during our simulations.

We have identified non-linear behaviors in the interplay among  opacities, which illustrate that the context provided by the  included opacities is important in evaluating individual opacities.
Some examples of neutrino opacity interplay include:
\begin{itemize}
  \item Emission from nuclear EC  and energy downscattering by NES compete to fill the  lowest energy bins  during collapse.
The escape of low-energy \nue\ increases core deleptonization.
The low-energy spectrum of neutrinos emitted by the LMSH EC table fills the low-energy phase space adequately without NES; thus, we do not see an impact on deleptonization when NES is omitted as we do in the case of models using the IPA EC, which does not fill this part of the spectrum directly.
  \item Thermalization of \numt\numtbar\ by individual NIS opacities is not simply additive.
  Removing either NES or nucleon NIS results in a modest increase in \meanE{\numt}, while removing both results in a much larger increase, as thermalization by scattering above the emission region is absent.
  \item Neutrino emission by pair sources also exhibits saturation effects.
In models including $e^+e^-$ annihilation, bremsstrahlung has only a minor impact on the emitted neutrino properties, but when bremsstrahlung is the only pair source, its removal has a much larger impact.
\end{itemize}

We can identify from our tested set necessary neutrino--matter interactions required for modern supernova modeling in any dimension.
\begin{itemize}
\item 
Modern nuclear EC \citep[LMSH, ][ or equivalent]{JuLaHi10} should be considered an essential ingredient in any realistic supernova simulation as was previously noted by \citet{HiMeMe03}.
Relying on the IPA EC artificially alters the electron capture, deleptonization, and the impact of other opacities.
\item Nucleon NIS extends the shock radius via an increase in \nue\nuebar-luminosity. The related enhancements to capture on free nucleons, though relatively modest in effect, should be included for physical consistency of the nucleon opacities.
\item NES  significantly reduces \nue\ emission during breakout and contributes to thermalizing the \numt\numtbar\ spectra.
\item \numt\numtbar-pair emission by $e^+e^-$-annihilation is an important source of cooling during the accretion phase, while bremsstrahlung plays only a small role \citep[as also seen in][]{ThBuPi03} unless $e^+e^-$-annihilation is omitted. Bremsstrahlung may become more important at later epochs as trapped \numt\numtbar\ diffuse out.
\end{itemize}
Omitting any of these opacities would alter the observable neutrino properties, and thus would introduce unnecessary systematic errors in the analysis of  observed supernova neutrino signals.

The modern opacities discussed in this paper are physically well-motivated improvements to the reference \citet{Brue85} opacity set.
Including these improved opacities increases the physical fidelity of the neutrino--matter interactions in supernova simulations, while omitting them  risks potential systematic errors in the dynamical and observational properties of simulated supernovae.

\acknowledgements

E.J.L. received support from the NASA Astrophysics Theory and Fundamental Physics Program (grant number NNH11AQ72I) and the NSF PetaApps Program (grant number OCI-0749242).
A.M. and W.R.H. are supported by the Department of Energy Office of Nuclear Physics; and A.M. and O.E.B.M. received support from  the Department of Energy Office of Advanced Scientific Computing Research. 
This research was also supported in part by the National Science Foundation through TeraGrid resources provided by National Institute for Computational Sciences under grant number TG-MCA08X010.
This research used resources of the Oak Ridge Leadership Computing Facility at the Oak Ridge National Laboratory, which is supported by the Office of Science of the U.S. Department of Energy under Contract No. DE-AC05-00OR22725.

\end{document}